\documentclass[11pt,a4paper]{article}
\usepackage{amsmath,amssymb,amsfonts}
\usepackage{epsfig}
\usepackage{axodraw4j}
\usepackage{color}
\usepackage[sort&compress,square]{natbib}
\usepackage[scale={0.8,0.8},vmarginratio=9:10,hmarginratio=1:1,headheight=20pt,footskip=40pt]{geometry}

\newcommand{\Aslash}{A\!\!\!/}
\newcommand{\Bslash}{B\!\!\!\!/}

\title{
\hfill{\small DCPT/09/20}\\[-0.2cm]\hfill{\small IPPP/09/10}\\[-0.2cm]\vspace{20pt} \Large{\textbf{
Minicharges and
Magnetic
Monopoles}}}
\author{Felix Br\"ummer\footnote{{\bf e-mail}: felix.bruemmer@durham.ac.uk}\,\, and Joerg Jaeckel\footnote{{\bf
e-mail}: joerg.jaeckel@durham.ac.uk}
\\[2ex]
\small{\it Institute for Particle Physics Phenomenology, Durham University, Durham DH1 3LE, UK}\\
}
\date{}

\begin{document}

\maketitle

\begin{abstract}
\noindent Minicharged particles arise naturally in extensions of the
Standard Model with a kinetic mixing term between the ordinary
electromagnetic U(1) and an extra ``hidden sector'' U(1). In this
note we study the compatibility of these particles with the
existence of magnetic monopoles. We find that angular momentum
quantization allows only certain combinations of ordinary and hidden
monopole charge. Using the example where one of the U(1)s originates
from a spontaneously broken SU(2), we demonstrate that exactly the
allowed types of monopoles arise as 't Hooft-Polyakov monopoles.
\end{abstract}

\vspace{3ex}

\section{Introduction}
Many extensions of the Standard Model contain additional U(1) gauge
factors. If the Standard Model particles are uncharged under an
extra U(1), it belongs to a so-called ``hidden sector'' and the
extra gauge boson could be light or even massless without violating
present experimental bounds. One of the most striking features of
theories with massless extra U(1) gauge bosons is that they
naturally lead to the appearance of particles with non-quantized
charges under the ordinary electromagnetic gauge
group~\cite{Holdom:1985ag}. These particles are often called
minicharged particles, since their charges are constrained by
experiment to be fairly small~\cite{Davidson:2000hf}.

On the other hand, if a U(1) gauge theory is to permit magnetic monopoles, charges must be quantized according to Dirac's quantization
condition~\cite{Dirac:1931kp,Dirac:1948um}\footnote{One should admit that so far neither monopoles nor minicharged particles have been found.}.
In purely U(1) theories the existence of monopoles is not necessary (and may even be problematic).
The issue becomes more pressing, however, if at least one of the U(1)s arises from a spontaneously broken compact non-abelian gauge group. In many such theories, monopoles of the 't Hooft-Polyakov type~\cite{'tHooft:1974qc,Polyakov:1974ek} are unavoidable. Nevertheless kinetic mixing and minicharged particles seem to be perfectly allowed from the point of view of the low-energy Lagrangian.
The main purpose of this note is to resolve this apparent contradiction.

To get started let us briefly define charge quantization somewhat more precisely. Charge quantization means that
all charges are integer multiples of a minimal charge, which in particular requires that
\begin{equation}
\label{quant}
\frac{q_{i}}{q_{j}}\in {\mathbb{Q}},
\end{equation}
for any two charges $q_{i}$ and $q_{j}$, where ${\mathbb{Q}}$ is
the field of rational numbers.

Let us now recall how particles with arbitrary irrational charges arise in theories with kinetic mixing.
The simplest model contains two U(1) gauge groups, one of which could {\it e.g.}~be our electromagnetic {U(1)$_{_\mathrm{QED}}$}, the other a hidden-sector {U(1)$_\mathrm{h}$} under
which all standard model particles have zero charge. In the gauge sector the most general
renormalizable Lagrangian allowed by the symmetries is
\begin{equation}
\label{lag}
{\mathcal{L}}=-\frac{1}{4} F^{\mu\nu}F_{\mu\nu}-\frac{1}{4}G^{\mu\nu}G_{\mu\nu}
-\frac{1}{2}\chi\,F^{\mu\nu}G_{\mu\nu},
\end{equation}
where $F_{\mu\nu}$ is the field strength tensor for the ordinary
electromagnetic {U(1)$_{_\mathrm{QED}}$} gauge field $A^{\mu}$, and
$G^{\mu\nu}$ is the field strength for the hidden-sector
{U(1)$_\mathrm{h}$} field $B^{\mu}$, {\it i.e.}, the hidden photon.  The
first two terms are the standard kinetic terms for the photon and
hidden photon fields, respectively. Because the field strength
itself is gauge invariant for U(1) gauge fields, the non-diagonal
third term is also allowed by gauge and Lorentz symmetry.  This is
the kinetic mixing term~\cite{Holdom:1985ag}.

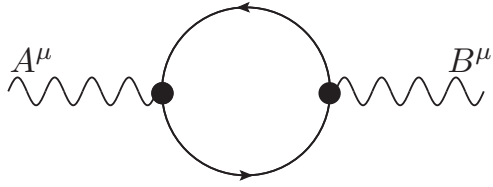
\begin{figure}
\begin{center}
\scalebox{0.7}[0.7]{\fcolorbox{white}{white}{
  \begin{picture}(258,92) (111,-82)
    \SetWidth{1.0}
    \SetColor{Black}
    \Arc[arrow,arrowpos=0.375,arrowlength=5,arrowwidth=2,arrowinset=0.2](240,-29)(45.255,135,495)
    \Arc[arrow,arrowpos=0.88,arrowlength=5,arrowwidth=2,arrowinset=0.2](240,-29)(45.255,135,495)
    \Photon(112,-29)(192,-29){7.5}{4}
    \Photon(288,-29)(368,-29){7.5}{4}
    \Vertex(195,-30){6}
    \Vertex(195,-30){6}
    \Vertex(285,-30){6}
    \Text(125,-12)[c]{\scalebox{1.7}[1.7]{$A^{\mu}$}}
    \Text(362,-12)[c]{\scalebox{1.7}[1.7]{$B^{\mu}$}}
  \end{picture}
  }}
\end{center}
\caption{Feynman diagram for a contribution of a particle charged under both gauge groups to the kinetic mixing.} \label{loopfig}
\end{figure}

From the viewpoint of a low-energy effective Lagrangian, $\chi$ is a
completely arbitrary parameter, which will be irrational in general.
Embedding our model into a more fundamental
theory, it is conceivable that $\chi=0$ holds at some UV-completion scale.
But even then, integrating out the
quantum fluctuations below this scale tends to generate a
non-vanishing~$\chi$~\cite{Holdom:1985ag}.
For example, integrating out a pair of heavy particles with charges
$(1,1)$ and $(1,-1)$ under the visible and
hidden sector gauge groups (with coupling strengths $e$ and $e_{\rm h}$
respectively),
we find at 1-loop order ({\it cf.} Fig.~\ref{loopfig})
\begin{equation}
\label{loopexp}
\chi=\frac{e e_{\rm h}}{6\pi^2}\log\left(\frac{m^{\prime}}{m}\right),
\end{equation}
where $m$ and $m^{\prime}$ are the masses of the two particles.
This is generically an irrational number.
In a similar manner, kinetic mixing arises in many
string theory models~\cite{Abel:2006qt,Abel:2008ai,Dienes:1996zr,Lust:2003ky,Abel:2003ue,Abel:2004rp,Batell:2005wa,Blumenhagen:2006ux}.

The kinetic term can be diagonalized by a shift
\begin{equation}
\label{shift}
B^{\mu}\rightarrow \tilde{B}^{\mu}-\chi A^{\mu}.
\end{equation}
Apart from a multiplicative renormalization of the gauge coupling,
\begin{equation}
\label{rescaling}
e^2\rightarrow e^2/(1-\chi^2),
\end{equation}
the visible-sector fields remain
unaffected by this shift. Consider now a hidden-sector
fermion\footnote{Here
  and in the following, we will specialize to the case where the
  hidden-sector particle is a fermion. A generalization to scalars is
  straightforward and does not change the results qualitatively.}  $h$
that has charge one under $B^{\mu}$. Applying the shift~\eqref{shift}
to the coupling term, we find:
\begin{equation}
e_{\rm h}\bar{h}\Bslash\, h\rightarrow e_{\rm h}\bar{h}\tilde\Bslash\, h-\chi e_{\rm h}\bar{h}\Aslash\,
 h.
\end{equation}
We can read off that the hidden-sector particle now has a charge
\begin{equation}
\label{epsiloncharge}
\epsilon e=-\chi e_{\rm h}
\end{equation}
under the visible electromagnetic gauge field $A^{\mu}$. Since $\chi$ is an
arbitrary number, the
fractional electric charge $\epsilon$ of the hidden-sector
fermion $h$ is not necessarily integer or rational.

One is often interested in the case $|\chi|\ll1$ and accordingly $|\epsilon|\ll 1$, because for masses below
the electroweak scale large electromagnetic couplings are ruled out by experiment and observation (see~\cite{Davidson:2000hf} and for some interesting
recent developments~\cite{Gies:2006ca,Badertscher:2006fm,Ehret:2007cm,Zavattini:2005tm}). This is why we will in the following
often refer to these particles as minicharged particles, although what we will say will also be valid for larger charges.

So despite the fact that we started with a particle carrying integral charge under the hidden sector gauge group, this particle now appears as a particle
with an arbitrary, not necessarily rational charge under the ordinary electromagnetic field.
Note, however, that a particle charged only under the ordinary electromagnetic gauge group will remain unaffected by the shift Eq.~\eqref{shift}.

\section{U(1) monopoles}\label{point}
Above we have written down an at first sight completely reasonable field theory that contains particles with not necessarily rational electric charges.
Let us now study what happens if we introduce monopoles. For simplicity we will mainly concentrate on a version of the charge quantization argument focussing on the field
angular momentum~\cite{Saha:1949,Wilson:1949}. The field angular momentum of a static configuration of an electric point charge with
charge $q e$ and a magnetic (point) monopole with charge $1\cdot g$ is given by
\begin{equation}
\mathbf{L}=\int d^{3}x\,\, \mathbf{x}\times(\mathbf{E}\times\mathbf{B})=\frac{qeg}{4\pi} \hat{\mathbf{n}}.
\end{equation}
Here, $\hat{\mathbf{n}}$ is the unit vector pointing from the electric charge to the magnetic charge, and the right hand side can be obtained
by inserting the electric fields $\mathbf{E}=qe\mathbf{r}/(4\pi r^3)$ and the magnetic field $\mathbf{B}=g\mathbf{r}/(4\pi r^3)$ for the electric and
magnetic monopole, respectively.
The quantization of angular momentum in quantum mechanics now requires
\begin{equation}
\label{anguquant}
|\mathbf{L}|=\frac{qeg}{4\pi}=\frac{n}{2},
\end{equation}
where $n$ is an integer (and as usual $\hbar=1$).
Requiring \eqref{anguquant} for all charges $q$ in the theory, we automatically enforce the quantization condition~\eqref{quant}.

Let us now see how this generalizes to our situation with two U(1) gauge fields. In the $(A^{\mu},\tilde{B}^{\mu})$-basis (or tilded basis) the kinetic terms
are diagonal and the generalized expression for the angular momentum is straightforward:
\begin{equation}
\label{angutilde}
\mathbf{L}=\int d^{3}x\,\, \mathbf{x}\times(\mathbf{E}\times\mathbf{B}+\tilde{\mathbf{E}}_{\rm h}\times\tilde{\mathbf{B}}_{\rm h}),
\end{equation}
where $\tilde{\mathbf{E}}_{\rm h}$ and $\tilde{\mathbf{B}}_{\rm h}$ are the hidden electric and magnetic fields in the tilded basis.

With two U(1) factors we can, of course, also have more general magnetic monopoles. In general a monopole can have charges
$(g,\tilde{g}_{\rm h})$ under the visible and hidden magnetic fields (in the tilded basis). Its magnetic field will then be,
\begin{equation}
\left(
  \begin{array}{l}
    \mathbf{B} \\
    \tilde{\mathbf{B}}_{\rm h} \\
  \end{array}
\right)
=\frac{\mathbf{r}}{4\pi r^3}
\left(
  \begin{array}{l}
    g \\
    \tilde{g}_{h} \\
  \end{array}
\right).
\end{equation}
We can now study static configurations of this monopole with:
\begin{itemize}
\item[a)]{} an ordinary electrically charged particle (charge $q=1$) with a field
\begin{equation}
\left(
  \begin{array}{c}
 \!\!   \mathbf{E} \\
    \tilde{\mathbf{E}}_{\rm h} \\
  \end{array}
\right)=\frac{\mathbf{r}}{4\pi r^3}\left(
                                \begin{array}{c}
                                  e \\
                                  0 \\
                                \end{array}
                              \right)
\end{equation}
and
\item[b)]{} a hidden sector particle (hidden charge $q_{\rm h}=1$) that has acquired
a charge $\epsilon$ under the ordinary electromagnetic field,
\begin{equation}
\left(
  \begin{array}{c}
 \!\!   \mathbf{E} \\
    \tilde{\mathbf{E}}_{\rm h} \\
  \end{array}
\right)=\frac{\mathbf{r}}{4\pi r^3}\left(
                                \begin{array}{c}
                                  \epsilon e \\
                                  e_{\rm h} \\
                                \end{array}
                              \right).
\end{equation}
\end{itemize}

Inserting into Eq.~\eqref{angutilde} we find
\begin{equation}
|\mathbf{L}_{a)}|=\frac{eg}{4\pi},\quad\quad|\mathbf{L}_{b)}|=\frac{\epsilon eg+e_{\rm h}\tilde{g}_{\rm h}}{4\pi}=\frac{-\chi e_{\rm h}g+e_{\rm h}\tilde{g}_{\rm h}}{4\pi}.
\end{equation}
Angular momentum quantization now requires that both configurations have half-integer angular momentum,
\begin{equation}
\label{constraint}
|\mathbf{L}_{a)}|=\frac{eg}{4\pi}=\frac{n}{2}\quad {\rm and}\quad |\mathbf{L}_{b)}|=\frac{\epsilon eg+e_{\rm h}\tilde{g}_{\rm h}}{4\pi}
=\frac{-\chi e_{\rm h}g+e_{\rm h}\tilde{g}_{\rm h}}{4\pi}=\frac{m}{2},
\end{equation}
where $m$ and $n$ are integers.
It is clear that a naive monopole with $\tilde{g}_{\rm h}=0$ causes a problem because this would require $|\mathbf{L}_{a)}|/|\mathbf{L}_{b)}|=\epsilon=n/m$
and therefore $\epsilon$ to be rational. In the previous section we have, however, argued that $\epsilon$ is typically not rational.
This is the apparent contradiction produced by introducing both monopoles and minicharged particles.

However, a closer inspection of Eq.~\eqref{constraint} reveals two types of monopoles which will not
cause any such problems for arbitrary $\chi$,
\begin{equation}
\label{allowed}
\left(
  \begin{array}{c}
    g \\
    \tilde{g}_{\rm h} \\
  \end{array}
\right)
=\frac{2\pi m}{e_{\rm h}}\left(
   \begin{array}{c}
     0 \\
     1 \\
   \end{array}
 \right),\quad{\rm and}\quad
\left(
  \begin{array}{c}
    g \\
    \tilde{g}_{\rm h} \\
  \end{array}
\right)
=\frac{2\pi n}{e}\left(
   \begin{array}{c}
     1 \\
     \chi \\
   \end{array}
 \right).
\end{equation}
For the first monopole $|\mathbf{L}_{a)}|=0$ and for the second $|\mathbf{L}_{b)}|=0$.

From the point of view of a true U(1) gauge theory which does not arise from a spontaneously broken non-abelian gauge theory, it is a priori not
clear what types of monopoles are allowed. Eq.~\eqref{allowed} then defines the types of monopole which may be consistently added to the theory.
As we will see in the next section, exactly these allowed types of monopoles appear as \mbox{'t Hooft-Polyakov} monopoles if (one of) the U(1)s arises from a spontaneously broken SU(2).

So far we have treated the problem in the tilded basis, where the kinetic term for the gauge fields is diagonal but the hidden sector particles
appear to be minicharged. Let us now see how the situation presents itself in the basis where the kinetic terms for the gauge fields are non-diagonal,
but all particles have integer charges. In this basis the electric fields for the particle with ordinary charge, a), and hidden charge, b), read
\begin{equation}
\label{efields}
a):\quad \left(
  \begin{array}{c}
 \!\!   \mathbf{E} \\
    \mathbf{E}_{\rm h} \\
  \end{array}
\right)=\frac{\mathbf{r}}{4\pi r^3}\left(
                                \begin{array}{c}
                                  e \\
                                  -\chi e \\
                                \end{array}
                              \right),
\quad{\rm and}\quad
b):\quad
\left(
  \begin{array}{c}
 \!\!   \mathbf{E} \\
    \mathbf{E}_{\rm h} \\
  \end{array}
\right)=\frac{\mathbf{r}}{4\pi r^3}\left(
                                \begin{array}{c}
                                  -\chi e_{\rm h} \\
                                  e_{\rm h} \\
                                \end{array}
                              \right).
\end{equation}

Before we can study the angular momentum of configurations with a monopole we have to briefly revisit the expression for the angular momentum. By undoing the gauge coupling rescaling of Eq.~\eqref{rescaling} and inverting the shift of Eq.~\eqref{shift}
using $\tilde{\mathbf{E}}_{\rm h}=\mathbf{E}_{\rm h}+\chi \mathbf{E}$ and $\tilde{\mathbf{B}}_{\rm h}=\mathbf{B}_{\rm h}+\chi \mathbf{B}$ (or by directly deriving the Poynting vector from the Lagrangian Eq.~\eqref{lag}), we find for the angular momentum in the original basis
\begin{equation}
\label{angu2}
{\mathbf{L}}=\int d^{3}x\,\mathbf{x}\times[\mathbf{E}\times\mathbf{B}+\chi(\mathbf{E}_{\rm h}\times\mathbf{B}
+\mathbf{E}\times\mathbf{B}_{\rm h})+\mathbf{E}_{\rm h}\times\mathbf{B}_{\rm h}].
\end{equation}

Inserting the fields for electric point charges, Eq.~\eqref{efields}, and for a magnetic monopole,
\begin{equation}
\left(
  \begin{array}{l}
    \mathbf{B} \\
    \mathbf{B}_{\rm h} \\
  \end{array}
\right)
=\frac{\mathbf{r}}{4\pi r^3}
\left(
  \begin{array}{l}
    g \\
    g_{h} \\
  \end{array}
\right),
\end{equation}
we find for the angular momentum quantization conditions
\begin{equation}
\label{constraint2}
|\mathbf{L}_{a)}|=\frac{eg}{4\pi}=\frac{n}{2}\quad {\rm and}\quad |\mathbf{L}_{b)}|=\frac{e_{\rm h}g_{\rm h}}{4\pi}=\frac{m}{2}.
\end{equation}
So in this basis it is exactly the `expected' monopoles with vanishing charge under one of the U(1)s, $g=0$ or $g_{\rm h}=0$,
which are allowed by the quantization conditions (of course, integer linear combinations of these are also allowed).

It is instructive to also briefly consider a charge quantization argument due to Goldhaber~\cite{Goldhaber:1965}. It focusses on
the scattering of an ordinary particle
on a monopole. For a monopole resting at the origin of the coordinate system, and the charged particle flying at a distance $b$
parallel to the $z$-axis, the Lorentz force is
\begin{equation}
\label{force}
a):\quad F_{y}=\frac{eg}{4\pi}\frac{vb}{(b^2+v^2t^2)^{\frac{3}{2}}},\quad\quad b):\quad
F_{y}=\frac{(\epsilon e g+\tilde{g}_{\rm h}e_{\rm h})}{4\pi}\frac{vb}{(b^2+v^2t^2)^{\frac{3}{2}}}
=\frac{g_{\rm h}e_{\rm h}}{4\pi}\frac{vb}{(b^2+v^2t^2)^{\frac{3}{2}}},
\end{equation}
where $v$ is the velocity of the charged particle and $t=0$ is the time when the particle is closest to the monopole.
Integrating the force over time we obtain the change in momentum, and more importantly angular momentum,
\begin{equation}
a):\quad \Delta L_{z}=\frac{eg}{2\pi},\quad\quad b):\quad \Delta L_{z}=\frac{(\epsilon e g+\tilde{g}_{\rm h}e_{\rm h})}{2\pi}=\frac{g_{\rm h}e_{\rm h}}{2\pi}.
\end{equation}
Assuming that angular momentum only changes by integer amounts, we are back at the quantization conditions Eqs.~\eqref{constraint} and \eqref{constraint2}.
Inserting, however, our allowed values for the monopole, it becomes clear that the allowed types of monopole are exactly those
where one of the electrically charged particles does not experience a change of angular momentum during the scattering. Moreover, looking at Eq.~\eqref{force}, it becomes
clear that it actually experiences no force at all. In this sense the ordinary monopoles are invisible to hidden electrically charged matter, and hidden monopoles
are invisible to ordinary electrically charged matter. This is in contrast to the situation between the electrically charged particles themselves:
Electrically charged ordinary particles do, albeit weakly, interact with the electrically charged hidden sector
particles\footnote{Similarly the two types of magnetic monopoles (weakly) interact with each other. This is in accord with the idea of electric-magnetic duality for the case of several
U(1)s, see for instance~\cite{Argyres}.}.

\section{'t Hooft-Polyakov monopoles}
In the previous section we have merely determined the types of monopoles which are allowed by angular momentum quantization. Our rationale was that in a theory with purely U(1) gauge groups the existence of monopoles is not forced upon us, and we can choose to introduce only the allowed ones.

On the other hand, it is well-known that some models necessarily contain magnetic monopoles. The most famous example is probably the 't Hooft-Polyakov monopole: An SU(2) gauge theory, when spontaneously broken to U(1) via an adjoint Higgs field, admits stable solitonic solutions to its classical field equations which represent magnetic monopoles. In a similar manner, monopoles may arise from other non-abelian gauge groups after spontaneous symmetry breaking. Their magnetic charges can be calculated and turn out to satisfy Dirac's quantization condition.

We will now show, using the 't Hooft-Polyakov model as an example, that the analogous statement holds even in the case of multiple U(1)s and kinetic mixing.
For simplicity we will consider a situation where only one of the U(1)s originates from an SU(2), but the generalization is straightforward.

Let us start with the Lagrangian for such an SU(2)$\times$U(1) theory,
\begin{equation}
\label{lagrangian}
{\mathcal L}=-\frac{1}{4}G^{a,\mu\nu}G^{a}_{\mu\nu}-\frac{1}{2}D^{\mu}Q^{a}D_{\mu}Q^{a}-V((Q^{a})^2)-\frac{1}{4}F^{\mu\nu}F_{\mu\nu}-\frac{1}{2M}Q^{a}G^{a,\mu\nu}F_{\mu\nu}.
\end{equation}
The first terms correspond to the usual kinetic term for an SU(2) gauge field and the kinetic and potential terms for an adjoint scalar
that breaks the SU(2) down to U(1). The second to last term is the kinetic term for the U(1) field. The last term is a gauge invariant term which, as we will see below,
results in kinetic mixing after spontaneous symmetry breaking.

For a suitable form of the potential $V$, the field $Q^{a}$ acquires a vev. We will turn to monopole solutions later, so for the moment the vev can be taken to be constant and to lie in the 3-direction,
\begin{equation}
\label{higgsvev}
\langle Q^{a}\rangle=(0,0,v).
\end{equation}
Then the $1$- and $2$-components of the gauge fields become massive and the $3$-component provides for the remaining U(1) field $B^{\mu}$ with
gauge field strength $G^{\mu\nu}$.
Using this identification, inserting the vev and retaining only the massless fields, Eq.~\eqref{lagrangian} then becomes equal to Eq.~\eqref{lag} with
\begin{equation}
\chi=\frac{v}{M}.
\end{equation}

Before we turn to consider monopoles, let us again stress that the last term in Eq.~\eqref{lagrangian} can indeed be generated naturally by integrating out
a heavy particle $\Psi$ coupled to both gauge groups,
\begin{equation}
{\mathcal L}_{\Psi}={\rm i}\bar{\Psi}_{i}\gamma^{\mu}(\partial_{\mu}\mathbf{1}_{ij}+ieA_{\mu}\mathbf{1}_{ij}+if W^{a}_{\mu}t^{a}_{ij})\Psi_{j}
-m_{0}\bar{\Psi}\Psi-hQ^{a}\bar{\Psi}_{i}t^{a}_{ij}\Psi_{j}.
\end{equation}
Here the $t^{a}=\sigma^{a}/2$ are the generators of SU(2), $e$ and $f$ are the gauge couplings of the U(1) and SU(2) fields $A_{\mu}$ and $W^{a}_{\mu}$, and $h$ is a Yukawa coupling
between the Higgs field $Q^{a}$ and $\Psi$.
In the above Higgs background, $\Psi$ acquires a mass matrix
\begin{equation}
\label{massmatrix}
m_{0}\left(
   \begin{array}{cc}
     1 & 0 \\
     0 & 1 \\
   \end{array}
 \right)
+\frac{hv}{2}\left(
    \begin{array}{cc}
      1 & 0 \\
      0 & -1 \\
    \end{array}
  \right).
\end{equation}
Due to the spontaneous breaking of the SU(2), the two components of $\Psi$ have different masses (this invalidates the naive
argument that the diagram shown in Fig.~\ref{loopfig} is $\sim {\rm Tr}(t^{a})=0$, because the $\Psi$ propagators are no longer proportional to
the unit matrix in SU(2) space). Therefore, we can indeed have non-vanishing kinetic mixing if the SU(2) is broken spontaneously.
Under the unbroken $W^{3}_{\mu}\equiv B_{\mu}$ field the first component, $\Psi_{1}$ has charge $1$ whereas the second component, $\Psi_{2}$,
has charge~$-1$. Hence the two
components of $\Psi$ provide for exactly the pair of particles charged under both expressions that lead to Eq.~\eqref{loopexp}
with the replacements $m=m_{0}+hv/2$ and and $m^{\prime}=m_{0}-hv/2$. Expanding to leading order in $hv$ and restoring the correct
index structure for gauge invariance, we recover exactly the last term in Eq.~\eqref{lagrangian} with
\begin{equation}
\frac{1}{M}=\frac{ef}{6\pi^2}\frac{h}{m_{0}}.
\end{equation}

Let us now turn to the monopole solutions. Following 't Hooft~\cite{'tHooft:1974qc} we parameterize a set of static and rotationally symmetric solutions for $Q^a$ and $W^a_\mu$ by
\begin{equation}
Q^{a}=r_{a} Q(r),\quad\quad W^{a}_{\mu}=\epsilon_{\mu a b} r_{b} W(r).
\end{equation}
Here $a,b,\ldots=1\ldots 3$ and $\epsilon_{\mu ab}$ is the usual epsilon symbol for $\mu=1\ldots 3$, while $\epsilon_{4 ab}=0$. The functions $W(r)$ and $Q(r)$ are determined by the equations of motion and by their asymptotic behaviour. We impose the boundary condition $Q(r)\rightarrow v/r$ at spacelike infinity, so the Higgs vev can, at large distances, locally always be gauge-transformed into the form of Eq.~\eqref{higgsvev}.

Using this ansatz, the part of the `effective' field strength pointing in the unbroken direction becomes
\begin{equation}
\label{Geff}
G_{{\rm eff},\mu\nu}=\frac{Q^{a}}{|Q|}G^{a}_{\mu\nu}=\frac{r_{a}}{r}\epsilon_{\mu\nu a}\,W(r)\left(2+f\,r^2 W(r)\right),
\end{equation}
where the epsilon symbol is again defined with $\epsilon_{4\nu a}=\epsilon_{\mu 4a}=0$, and $|Q|=|Q^a|=r\,Q(r)$. This corresponds to an effective magnetic field,
\begin{equation}
\label{effective}
\mathbf{B}_{\rm eff}=-\frac{\mathbf{r}}{r}\,W(r)\left(2+f\,r^2 W(r)\right).
\end{equation}
As one might expect, the effective electric field vanishes, $\mathbf{E}_{\rm eff}=0$.

In the absence of a kinetic mixing term, the large-distance behavior of $W(r)$ can be determined to be
\begin{equation}
\label{asympt}
W(r)\sim -\frac{1}{f r^2},
\end{equation}
and the field looks like that of a magnetic monopole,
\begin{equation}
\label{asympt2}
\mathbf{B}_{\rm eff}=g\frac{\mathbf{r}}{4\pi r^3},\quad g=\frac{4\pi}{f},
\end{equation}
automatically fulfilling the charge quantization conditions.

The crucial question is now if the monopole solution is affected by the presence of the kinetic mixing term (the last term in Eq.~\eqref{lagrangian}). In particular, will there be a non-vanishing $F_{\mu\nu}$?
We will argue that $F_{\mu\nu}=0$ is a consistent solution of the equations of motion and that the monopole solution is completely unchanged by the
kinetic mixing term.

In the equations of motion for $W^{a}_{\mu}$ and $Q^{a}$, the contributions from the last term in Eq.~\eqref{lagrangian} is clearly $\sim F_{\mu\nu}$.
Therefore, if $F_{\mu\nu}=0$ the extra term in these equations vanishes, and the same solutions as in the case without kinetic mixing solve these equations.

The only remaining equation is the one for $F_{\mu\nu}$,
\begin{equation}
\label{Fmunu}
\partial_{\mu}F^{\mu\nu}=-\frac{1}{M}\partial_{\mu}(Q^{a}G^{a,\mu\nu}).
\end{equation}
 By Eq.~\eqref{Geff}, the right-hand side of Eq.~\eqref{Fmunu} has the form
\begin{equation}
 -\frac{1}{M}\partial_{\mu}(Q^{a}G^{a,\mu\nu})=\partial_{\mu}\epsilon_{\mu\nu a} r_a\,F(r)
\end{equation}
with some regular function $F(r)$. Such an expression is zero by symmetry. Therefore Eq.~\eqref{Fmunu} reduces to the
sourceless Maxwell equations, for which  $F_{\mu\nu}=0$, $\mathbf{E}=\mathbf{B}=0$, is of course a consistent solution.

Let us now identify the unbroken U(1) of the SU(2) with our hidden sector gauge group, with $e_{\rm h}=f$. It is then suggestive to use $G_{{\rm eff}, \mu\nu}=G_{\mu\nu}$.
We obtain a Lagrangian of the form \eqref{lag} with an effective mixing parameter $\chi=|Q|/M$.
Using $\mathbf{B}=0$, the asymptotic solution Eqs.~\eqref{asympt}, \eqref{asympt2} represents a
monopole with $g=0$ and $g_{\rm h}=1/e_{\rm h}$. This is to be compared with the result obtained in Sect.~\ref{point} for the
unshifted basis: we find that this is exactly the allowed type of monopole.

\section{Conclusions}
If two U(1)s are coupled via a kinetic mixing term, a particle with integer charge under one of the U(1)s may appear to have an irrational charge proportional to the kinetic mixing under the other U(1). If also magnetic monopoles are introduced, such arbitrary `minicharges' lead to an
apparent conflict with the quantization of angular momentum. In this note we have shown that this problem is resolved
if the magnetic monopoles have appropriate monopole charges under both U(1)s. Indeed the solution is quite simple, the allowed monopoles
are integer linear combinations of those monopoles that do not interact with particles charged under one of the U(1)s.

In a purely U(1) setup we can include suitable consistent monopoles at will. However, in models where one or both of the U(1) arise from a spontaneously
broken non-abelian gauge group such as SU(2), monopoles can be automatically present in the spectrum, with their properties completely determined by the model.
Using a U(1)$\times$SU(2) theory broken to U(1)$\times$U(1) as an example, we have explicitly demonstrated how a kinetic mixing term
can arise upon spontaneous symmetry breaking. We have furthermore shown that it leaves the monopole solutions unaffected and gives exactly the allowed type
of monopoles.

In conclusion, in theories with kinetic mixing, minicharged particles are consistent with the existence of magnetic monopoles. Therefore, searches
for minicharged particles~\cite{Gies:2006ca,Badertscher:2006fm,Ehret:2007cm,Zavattini:2005tm} can also probe extra U(1) factors arising from non-abelian gauge groups.
This gives us one more opportunity to get an experimental glimpse on hidden sectors.

\section*{Acknowledgements}
We would like to thank C.~Durnford and V.~V.~Khoze for interesting discussions and useful comments.


\begin{thebibliography}{10}
\bibitem{Holdom:1985ag}
  B.~Holdom,
  Phys.\ Lett.\  B {\bf 166}, 196 (1986).

\bibitem{Davidson:2000hf}
  S.~Davidson, S.~Hannestad and G.~Raffelt,
  JHEP {\bf 0005} (2000) 003
  [arXiv:hep-ph/0001179].

\bibitem{Dirac:1931kp}
  P.~A.~M.~Dirac,
  Proc.\ Roy.\ Soc.\ Lond.\  A {\bf 133} (1931) 60.

\bibitem{Dirac:1948um}
  P.~A.~M.~Dirac,
  Phys.\ Rev.\  {\bf 74} (1948) 817.

\bibitem{'tHooft:1974qc}
  G.~'t Hooft,
  Nucl.\ Phys.\  B {\bf 79} (1974) 276.


\bibitem{Polyakov:1974ek}
  A.~M.~Polyakov,
  JETP Lett.\  {\bf 20} (1974) 194
  [Pisma Zh.\ Eksp.\ Teor.\ Fiz.\  {\bf 20} (1974) 430].


\bibitem{Abel:2006qt}
  S.~A.~Abel, J.~Jaeckel, V.~V.~Khoze and A.~Ringwald,
  Phys.\ Lett.\  B {\bf 666} (2008) 66
  [arXiv:hep-ph/0608248].

\bibitem{Abel:2008ai}
  S.~A.~Abel, M.~D.~Goodsell, J.~Jaeckel, V.~V.~Khoze and A.~Ringwald,
  JHEP {\bf 0807} (2008) 124
  [arXiv:0803.1449 [hep-ph]].



\bibitem{Dienes:1996zr}
  K.~R.~Dienes, C.~F.~Kolda and J.~March-Russell,
  Nucl.\ Phys.\  B {\bf 492} (1997) 104
  [hep-ph/9610479].

\bibitem{Lust:2003ky}
  D.~Lust and S.~Stieberger,
  Fortsch.\ Phys.\  {\bf 55} (2007) 427
  [hep-th/0302221].


\bibitem{Abel:2003ue}
  S.~A.~Abel and B.~W.~Schofield,
  Nucl.\ Phys.\  B {\bf 685}, 150 (2004)
  [hep-th/0311051].

\bibitem{Abel:2004rp}
  S.~Abel and J.~Santiago,
  J.\ Phys.\ G {\bf 30}, R83 (2004)
  [hep-ph/0404237].



\bibitem{Batell:2005wa}
  B.~Batell and T.~Gherghetta,
  Phys.\ Rev.\  D {\bf 73} (2006) 045016
  [arXiv:hep-ph/0512356].

\bibitem{Blumenhagen:2006ux}
  R.~Blumenhagen, S.~Moster and T.~Weigand,
  Nucl.\ Phys.\  B {\bf 751} (2006) 186
  [arXiv:hep-th/0603015].

\bibitem{Gies:2006ca}
  H.~Gies, J.~Jaeckel and A.~Ringwald,
  Phys.\ Rev.\ Lett.\  {\bf 97} (2006) 140402
  [hep-ph/0607118];
  M.~Ahlers, H.~Gies, J.~Jaeckel and A.~Ringwald,
  Phys.\ Rev.\  D {\bf 75} (2007) 035011
  [hep-ph/0612098];
  M.~Ahlers, H.~Gies, J.~Jaeckel, J.~Redondo and A.~Ringwald,
  Phys.\ Rev.\  D {\bf 77} (2008) 095001
  [0711.4991 [hep-ph]];
  M.~Ahlers, J.~Jaeckel and A.~Ringwald,
  arXiv:0812.3150 [hep-ph].

\bibitem{Ehret:2007cm}
  K.~Ehret {\it et al.},
  arXiv:hep-ex/0702023;
  C.~Robilliard,{\it et al}
  [BMV Collaboration],
  Phys.\ Rev.\ Lett.\  {\bf 99} (2007) 190403
  [arXiv:0707.1296 [hep-ex]];
  A.~S.~Chou {\it et al.}  [GammeV (T-969) Collaboration],
  Phys.\ Rev.\ Lett.\  {\bf 100} (2008) 080402
  [arXiv:0710.3783 [hep-ex]];
  P.~Pugnat {\it et al.}  [OSQAR Collaboration],
  arXiv:0712.3362 [hep-ex].
  M.~Fouche {\it et al.} [BMV Collaboration],
  Phys.\ Rev.\  D {\bf 78} (2008) 032013
  [arXiv:0808.2800 [hep-ex]];
  A.~Afanasev {\it et al.},
  Phys.\ Rev.\ Lett.\  {\bf 101} (2008) 120401
  [arXiv:0806.2631 [hep-ex]].


\bibitem{Zavattini:2005tm}
  E.~Zavattini {\it et al.}  [PVLAS Collaboration],
  Phys.\ Rev.\ Lett.\  {\bf 96} (2006) 110406
  [Erratum-ibid.\  {\bf 99} (2007) 129901]
  [arXiv:hep-ex/0507107];
  S.~J.~Chen, H.~H.~Mei and W.~T.~Ni [Q\& A Collaboration],
  Mod.\ Phys.\ Lett.\  A {\bf 22} (2007) 2815
  [arXiv:hep-ex/0611050];
P.~Pugnat {\em et al.} [OSQAR Collaboration],
Czech.\ J.\ Phys.\  {\bf 55}, A389 (2005);
Czech.\ J.\ Phys.\  {\bf 56}, C193 (2006);
R.~Battesti {\em et al.} [BMV Collaboration], Eur.\ Phys.\ J.\ D\ {\bf 46}, 323–333 (2008);
  E.~Zavattini {\it et al.}  [PVLAS Collaboration],
  Phys.\ Rev.\  D {\bf 77} (2008) 032006.
  [arXiv:0706.3419 [hep-ex]];
  G.~Cantatore, R.~Cimino, M.~Karuza, V.~Lozza and G.~Raiteri,
  arXiv:0809.4208 [hep-ex].

\bibitem{Badertscher:2006fm}
  A.~Badertscher {\it et al.},
  Phys.\ Rev.\  D {\bf 75} (2007) 032004
  [hep-ex/0609059];
  S.~N.~Gninenko, N.~V.~Krasnikov and A.~Rubbia,
  Phys.\ Rev.\  D {\bf 75} (2007) 075014
  [hep-ph/0612203].

\bibitem{Wilson:1949}
  H.~A.~Wilson,
  Phys.\ Rev.\  {\bf 75} (1949) 309.

\bibitem{Saha:1949}
  M.~N.~Saha, Indian\ J.\ Phys. {\bf 10} (1936) 145;
  Phys.\ Rev.\  {\bf 75} (1949) 1968.

\bibitem{Goldhaber:1965}
  A.~S.~Goldhaber,
  Phys.\ Rev.\  {\bf 140} (1965) B1407.

\bibitem{Argyres}
  P.~C.~Argyres,
{\it Prepared for ICTP Spring School on Nonperturbative Aspects of String Theory and Supersymmetric Gauge Theories, Trieste, Italy, 23-31 Mar
1998}

\end{thebibliography}
\end{document}